\documentclass[11pt]{llncs}
\usepackage{graphicx}
\usepackage{cite}
\usepackage{url}
\usepackage{times}
\usepackage{mathfont}
\def\min{\mathop{\rm min}}
\def\max{\mathop{\rm max}}

\def\inf{\mathop{\rm inf}}

\def\argmax{\mathop{{\rm arg\,max}}}

\mathcode`O="724F

\DeclareSymbolFont{AMSb}{U}{msb}{m}{n}
\DeclareSymbolFontAlphabet{\Bbb}{AMSb}
\def\R{\ensuremath{\Bbb R}}
\def\Z{\ensuremath{\Bbb Z}}

\DeclareMathSymbol{\subsetneq}{\mathrel}{AMSb}{"28}

\makeatletter
\def\hb@xt@{\hbox to }
\makeatother

\let\oldendproof\endproof
\def\endproof{\qed\oldendproof}

\begin{document}

\title{Quasiconvex Analysis of Backtracking Algorithms}
\author{David Eppstein}
\authorrunning{Eppstein}
\institute{School of Information \& Computer Science\\
University of California, Irvine\\
Irvine, CA 92697-3425, USA\\
\email{eppstein@ics.uci.edu}}

\date{ }

\maketitle
\begin{abstract}
We consider a class of multivariate recurrences frequently arising in the worst case analysis of Davis-Putnam-style exponential time backtracking algorithms for NP-hard problems.  We describe a technique for proving asymptotic upper bounds on these recurrences, by using a suitable weight function to reduce the problem to that of solving univariate linear recurrences; show how to use quasiconvex programming to determine the weight function yielding the smallest upper bound; and prove that the resulting upper bounds are within a polynomial factor of the true asymptotics of the recurrence.  We develop and implement a multiple-gradient descent algorithm for the resulting quasiconvex programs, using a real-number arithmetic package for guaranteed accuracy of the computed worst case time bounds.
\end{abstract}

\section{Introduction}

The topic of exponential-time exact algorithms for hard problems has led to much research in recent years \cite{Bei-SODA-99,Bys-SODA-03,DanHir-TR-00,Epp-SODA-01,Epp-WADS-01,csds0302030,GraHirNie-SAT-00,PatPudSak-FOCS-98,Sch-FOCS-99}.  In contrast to the situation with polynomial time algorithms, one can not significantly increase the size of a solvable instance by waiting for Moore's law to provide faster computers, so algorithmic improvement is especially important in this area.  Several design principles are known for such algorithms, including dynamic programming~\cite{Rob-Algs-86} and randomized hill climbing~\cite{Sch-FOCS-99}, but the most common approach is a simple form of branch-and-bound in which one repeatedly performs some case analysis to find an appropriate structure in the problem instance, and then uses that structure to split the problem into several smaller subproblems which are solved by recursive calls to the algorithm.

As an example of this approach, a graph coloring algorithm of the author~\cite{Epp-WADS-01} uses a subroutine for listing each maximal independent set of at most $k$ vertices ($k$-MIS) in an $n$-vertex graph.  This subroutine (slightly simplified for this example)
repeatedly selects and applies one of the following cases:
\begin{itemize}
\item If the input graph $G$ contains a vertex $v$ of degree zero,
recursively list each $(k-1)$-MIS in $G\setminus\{v\}$
and append $v$ to each listed set.
\item If the input graph $G$ contains a vertex $v$ of degree one, 
with neighbor $u$, recursively list each $(k-1)$-MIS in $G\setminus N(u)$ and append $u$ to each listed set.  Then, recursively list each $(k-1)$-MIS in $G\setminus \{u,v\}$ and append $v$ to each listed set.
\item If the input graph $G$ contains a path $v_1$-$v_2$-$v_3$ of degree-two vertices,
then, first, recursively list each $(k-1)$-MIS in $G\setminus N(v_1)$ and append $v_1$ to each listed set.  Second, list each $(k-1)$-MIS in $G\setminus N(v_2)$ and append $v_2$ to each listed set.  Finally, list each $(k-1)$-MIS in $G\setminus(\{v_1\}\cup N(v_3))$ and append $v_3$ to each listed set.  Note that, in the last recursive call, $v_1$ may belong to $N(v_3)$ in which case
the number of vertices is only reduced by three.
\item If the input graph $G$ contains a vertex $v$ of degree three or more,
recursively list each $k$-MIS in $G\setminus\{v\}$.
Then, recursively list each $(k-1)$-MIS in
$G\setminus N(v)$ and append $v$ to each listed set.
\end{itemize}
It is not hard to see that any graph contains at least one of these cases.
We can bound the worst-case number of output sets,
as a recurrence in the variables $n$ and $k$:
$$
T(n,k)=\max\left\{
\begin{array}{l}
T(n-1,k-1)\\
2T(n-2,k-1)\\
3T(n-3,k-1)\\
T(n-1,k)+T(n-4,k-1)
\end{array}
\right.
$$
As base cases, $T(0,0)=1$, $T(n,-1)=0$, and $T(n,k)=0$ for $k>n$.
Each term in the overall maximization of the recurrence comes from a case in the case analysis; the recurrence uses the maximum of these terms because, in a worst-case analysis, the algorithm has no control over which case will arise.  Each summand in each term comes from a recursive subproblem called for that case.  It turns out that, for the range of parameters of interest $n/4\le k\le n/3$, the recurrence above is dominated by its last two terms, and has the solution $T(n,k)=(4/3)^n(3^4/4^3)^k$.  We can also find graphs having this many $k$-MISs, so the analysis given by the recurrence is tight.
Similar but somewhat more complicated multivariate recurrences have arisen in our algorithm for
3-coloring~\cite{Epp-SODA-01} with variables counting 3- and 4-value variables in a constraint satisfaction instance, and in our algorithm for the traveling salesman problem in cubic graphs~\cite{csds0302030} with variables counting vertices, unforced edges, forced edges, and 4-cycles of unforced edges.

These examples of recurrences have all had few enough terms that they can be solved by hand, but Table~\ref{tbl:bigrec} depicts a recurrence, arising from unpublished work with J. Byskov on graph coloring algorithms, that is complex enough that hand solution seems unlikely.
This recurrence was derived through an iterative process, starting from a simple case analysis of the problem, in which the worst cases for the algorithm were repeatedly identified and replaced by a larger number of better cases.

\begin{table}
{\tiny $$
T(n,h) \le\max\left\{
\begin{array}{l}
T(n+3,h-2)+T(n+3,h-1)+T(n+4,h-2)+T(n+5,h-2) , \\
T(n,h+1)+T(n+1,h+2) , \\
2\,T(n+2,h)+2\,T(n+3,h) , \\
2\,T(n+2,h)+2\,T(n+3,h) , \\
T(n+3,h-2)+T(n+3,h-1)+T(n+5,h-3)+T(n+5,h-2) , \\
T(n+1,h)+T(n+3,h-1)+3\,T(n+3,h+3) , \\
T(n+3,h-2)+2\,T(n+3,h-1)+T(n+7,h-2) , \\
T(n+1,h)+2\,T(n+4,h-2) , \\
3\,T(n+1,h+2)+2\,T(n+1,h+5) , \\
2\,T(n+2,h)+T(n+3,h+1)+T(n+4,h)+T(n+4,h+1) , \\
T(n+1,h-1)+T(n+4,h-1) , \\
T(n+1,h+3)+2\,T(n+2,h)+T(n+3,h) , \\
2\,T(n+2,h-1) , \\
T(n,h+3)+T(n+1,h+2)+T(n+2,h) , \\
T(n+1,h-1)+T(n+4,h-1) , \\
2\,T(n+1,h+1)+T(n+2,h+1) , \\
9\,T(n+2,h+3) , \\
T(n+1,h)+T(n+1,h+1) , \\
9\,T(n+9,h-5)+9\,T(n+9,h-4) , \\
T(n+3,h-2)+T(n+3,h-1)+T(n+5,h-2)+2\,T(n+6,h-3) , \\
T(n+1,h-1)+T(n+4,h)+T(n+4,h+1) , \\
2\,T(n+2,h)+T(n+3,h)+T(n+4,h)+T(n+5,h) , \\
T(n+1,h)+2\,T(n+2,h+1) , \\
T(n+1,h-1) , \\
2\,T(n+2,h+1)+T(n+3,h-2)+T(n+3,h) , \\
T(n+1,h+1)+T(n+1,h+2)+T(n+2,h) , \\
2\,T(n+2,h)+2\,T(n+3,h) , \\
T(n+1,h+2)+T(n+2,h-1)+T(n+2,h+1) , \\
T(n+1,h) , \\
T(n+2,h+1)+T(n+3,h-2)+T(n+4,h-3) , \\
T(n-1,h+2) , \\
3\,T(n+4,h)+7\,T(n+4,h+1) , \\
T(n+2,h-1)+2\,T(n+3,h-1) , \\
T(n+2,h-1)+T(n+2,h)+T(n+2,h+1) , \\
T(n+3,h-2)+T(n+3,h)+2\,T(n+4,h-2) , \\
T(n+1,h)+T(n+3,h-1)+T(n+3,h+3)+T(n+5,h)+T(n+6,h-1) , \\
2\,T(n+1,h+4)+3\,T(n+3,h+1)+3\,T(n+3,h+2) , \\
3\,T(n+3,h+1)+T(n+3,h+2)+3\,T(n+3,h+3)+3\,T(n+4,h) , \\
T(n+2,h-1)+T(n+3,h-1)+T(n+4,h-2) , \\
T(n,h+1) , \\
T(n+1,h+2)+T(n+3,h-2)+T(n+3,h-1) , \\
2\,T(n+3,h-1)+T(n+3,h+2)+T(n+5,h-2)+T(n+5,h-1)+T(n+5,h)+2\,T(n+7,h-3) , \\
T(n+2,h+2)+2\,T(n+3,h)+3\,T(n+3,h+1)+T(n+4,h) , \\
T(n+3,h-2)+T(n+3,h-1)+T(n+5,h-3)+T(n+6,h-3)+T(n+7,h-4) , \\
T(n+1,h-1) , \\
T(n+1,h)+2\,T(n+3,h) , \\
4\,T(n+3,h+1)+5\,T(n+3,h+2) , \\
4\,T(n+2,h+3)+3\,T(n+4,h)+3\,T(n+4,h+1) , \\
T(n+3,h-2)+2\,T(n+3,h-1)+T(n+6,h-3) , \\
4\,T(n+2,h+3)+6\,T(n+3,h+2) , \\
T(n,h+1)+T(n+4,h-3) , \\
T(n+1,h-1)+2\,T(n+3,h+2) , \\
2\,T(n+2,h+1)+3\,T(n+2,h+3)+2\,T(n+2,h+4) , \\
2\,T(n+2,h)+2\,T(n+2,h+3) , \\
2\,T(n+2,h)+T(n+2,h+3)+T(n+3,h+2)+T(n+4,h)+T(n+4,h+1) , \\
2\,T(n,h+2) , \\
T(n+2,h)+T(n+3,h-2)+T(n+3,h-1) , \\
T(n+3,h-2)+2\,T(n+4,h-2)+T(n+5,h-3) , \\
T(n+1,h)+T(n+5,h-4)+T(n+5,h-3) , \\
T(n+1,h+2)+T(n+2,h-1)+T(n+3,h-1) , \\
T(n+2,h-1)+T(n+2,h)+T(n+4,h-1) , \\
10\,T(n+3,h+2) , \\
6\,T(n+2,h+2) , \\
T(n+2,h)+T(n+3,h) , \\
2\,T(n+3,h-1)+T(n+3,h+2)+T(n+5,h-2)+T(n+5,h-1)+T(n+5,h)+T(n+6,h-2)+T(n+7,h-2) , \\
6\,T(n+3,h+1) , \\
3\,T(n,h+3) , \\
T(n+2,h-1)+T(n+2,h)+T(n+4,h-2) , \\
2\,T(n+5,h-3)+5\,T(n+5,h-2) , \\
2\,T(n+2,h)+T(n+2,h+1)+T(n+4,h-1) , \\
8\,T(n+1,h+4) , \\
T(n+3,h-2)+T(n+3,h-1)+T(n+5,h-3)+T(n+5,h-2)+T(n+7,h-3) , \\
T(n+1,h-1)+T(n+2,h+2) , \\
5\,T(n+2,h+2)+2\,T(n+2,h+3)
\end{array}
\right.

$$}
\medskip
\caption{A recurrence arising from unpublished work with J. Byskov on graph coloring algorithms.}
\label{tbl:bigrec}
\end{table}

Our interest in this paper is in performing this type of analysis algorithmically: if we are given as input a recurrence such as the ones discussed above, can we efficiently determine its asymptotic solution, and determine which of the cases in the analysis are the critical ones for the performance of the backtracking algorithm that generated the recurrence?  We show that the answer is yes, by expressing the problem as a {\em quasiconvex program}, a type of generalized linear program studied previously by the author and others with applications including finite element mesh smoothing~\cite{AmeBerEpp-Algs-99}, brain flat mapping, hyperbolic graph drawing, and conformal meshing~\cite{BerEpp-WADS-01-omt}, and multi-projector tiled display color gamut equalization~\cite{BerEpp-SCG-03}.  This quasiconvex programming formulation allows us to solve a $\tau$-term recurrence in $O(\tau)$ steps, each of which involves the solution of a constant number of algebraic equations; alternatively, we can apply any of several numerical hill-climbing techniques, which are guaranteed to converge to the global optimum of a quasiconvex program.  We describe one such technique and provide two proof-of-concept implementations of it, one for exploratory analysis using floating point, and another using the {\tt XR} exact real-number computation package for Python~\cite{Python,XR}.  The algorithms we describe are able to analyze recurrences such as the one in Table~\ref{tbl:bigrec}, producing asymptotic analysis of their behavior (here, we are interested in the asymptotics of $T(n,0)$) and identifying the cases that are bottlenecks for that analysis,
typically within one or two seconds for a floating point evaluation sufficiently accurate for exploratory analysis of algorithms.

We first began using a weighting technique for upper bounding recurrences of this type in our previous work on graph coloring~\cite{Epp-SODA-01}, without much regard to its completeness or generality until this paper.
Despite some searching we have been unable to find relevant prior research on asymptotic analysis of similar multidimensional recurrences.

\section{Formalization and Statement of Results}

We assume that the input to our problem consists of the following items:
\begin{itemize}
\item An integer dimension $d$.
\item A recurrence
$$F(x) = \max_i \sum_j F(x - \delta_{i,j}),$$
where $x$ and $\delta_{i,j}$ are vectors in $\Z^d$, and $i$ and $j$ are indices ranging over the cases and subproblems of the algorithm to be analyzed.  We assume that, as base cases for the recurrence, $F(0)=1$, and $F(y)=0$ when $y$ can not reach the zero vector by any expansion of the recurrence.
\item A target vector $t$ in $\Z^d$.
\end{itemize}
The desired output is a description of the asymptotic behavior of the function
$f(n)=F(n\,t)$.  
We call the expressions $\sum_j F(x-\delta_{i,j})$ {\em terms} of the recurrence,
and their subexpressions $F(x-\delta_{i,j})$  {\em summands} of the term.
Much of what we describe here would generalize without difficulty to non-integer
values of $x$ and $\delta_{i,j}$,  and to non-integer multipliers on the summands of each term.

For our Python implementation, we represent the recurrence as a dictionary mapping case names to terms, with each term represented as a list of summands and each summand represented as a $d$-tuple of integers.  For instance, the cardinality-bounded maximal independent set recurrence discussed in the introduction has the representation shown in Table~\ref{tbl:repn}.
This Python representation also allows the inclusion of comments in the file containing the recurrence, describing in further detail the case analysis from which the recurrence arises.

\begin{table}[t]
\begin{verbatim}smallmis = {
    "deg0": [(1,1)],
    "deg1": 2*[(2,1)],
    "deg2": 3*[(3,1)],
    "deg3": [(4,1), (1,0)],
}\end{verbatim}
\caption{Python representation of recurrence input.}
\label{tbl:repn}
\end{table}

We obtain the following results:

\begin{itemize}
\item We show that our recurrences can be upper bounded by linear univariate recurrences formed by weighting the variables, and that the optimal set of weights for this upper bound technique can be found efficiently using quasiconvex programming.
The optimal basis for the quasiconvex program consists of the terms forming the worst cases for the recurrence. 
\item We describe a numerical improvement technique for quasiconvex programming, based on multi-gradient descent, and provide two implementations of this technique: one based on floating point arithmetic and capable of running at interactive speeds for exploratory algorithm analysis, and one using an exact arithmetic package for publishable guaranteed worst case bounds.
\item We prove lower bounds showing that the upper bounds from our optimal weighting technique are tight to within a polynomial factor.
\end{itemize}

\section{Upper Bounds}

In this section we describe a weighting technique that can be used to obtain asymptotic upper bounds for our recurrences, and formulate the problem of optimizing the weights in order to obtain the best such upper bound.

Fix a vector $w\in\R^d$, such that, for each summand $F(x-\delta_{i,j})$
of the input recurrence, $w\cdot\delta_{i,j}$ is positive, let $y$ be a real variable,
and define
$$F_w(y)=\max_{w\cdot x\le y} F(x).$$
Then it is not hard to see, by using the recurrence for $F(x)$ to expand the right hand side of this definition, and  interchanging the order in the maximization, that
$$F_w(y)\le\max_i\sum_j F_w(y-w\cdot\delta_{i,j}).$$
As base cases set $F_w(y)=1$ for $0\le y<\min_{i,j} w\cdot\delta_{i,j}$ and $F_w(y)=0$ for $y<0$.
This resembles the recurrence defining $F$, but with a real instead of vector-valued argument.   For linear univariate recurrences similar to this one,
standard solution techniques such as generating functions and characteristic polynomials are commonly taught in freshman combinatorics courses.
The recurrence for $F_w$ is nonlinear because of the maximization, and involves non-integer variables, but the same techniques apply:

\begin{lemma}
$F_w(y)=O(c_w^y)$, where $c_w$ is the unique positive root of the monotonic
function
$$r_w(c)=1-\max_i\sum_j c^{-w\cdot\delta_{i,j}}.$$
\end{lemma}

It will be convenient later to separate out this analysis into the different terms of the recurrence:

\begin{lemma}
Let $c_{w,i}$ denote the unique positive root of the monotonic function
$r_{w,i}(c)=1-\sum_j c^{-w\cdot\delta_{i,j}}$.
Then $c_w=\max_i c_{w,i}$.
\end{lemma}

Since $F_w$ was defined as a maximum of values of $F$, this technique immediately leads to upper bounds for the target function
$f(n)=F(n\, t)$:

\begin{lemma}\label{lem:wup}
Let $w\in\R^d$ be such that, for each summand $F(x-\delta_{i,j})$
of the input recurrence, $w\cdot\delta_{i,j}$ is positive, and let $w\cdot t=1$.
Then $f(n)\le F_w(n)=O(c_w^n)$.
\end{lemma}

We call $w$ a {\em weight vector}, because if one interprets the coordinates of $w$ as weights of the different backtracking algorithm instance features counted by $x$, then $w\cdot x$ is the total weight of the instance.  The recurrence for $F_w$ and its solution $O(c_w^n)$ describes the time for the backtracking algorithm as a function of instance weight.  However, different choices of the weight vector will give different upper bounds  $O(c_w^n)$ on the time used by the backtracking algorithm.  Our task now is to select the best weight vector, that is, the one yielding the tightest upper bound.  For convenience, we define $c_{w}=c_{w,i}=+\infty$ when
$w\cdot\delta_{i,j}$ is non-positive for some~$j$.

\section{Quasiconvex Programming}

In this section we find efficient algorithmic solutions for the optimal weighting problem  formulated in the previous section.

A function $q(w):\R^d\mapsto\R$ is called {\em quasiconvex} when its level sets
$q^{\le\lambda}=\{w\in R^d\mid q(x)\le\lambda\}$ are all convex.  In particular, the points $w$ where $q$ achieves its minimum value (if a minimum exists) form a convex set, and an approximation to a value achieving the global minimum can be found numerically by local improvement techniques.  If the functions $q_i$ for $i$ in some finite index set $S$ are all quasiconvex, then the function $q_S(w)=\max_{i\in S} q_i(w)$ is also quasiconvex, and it becomes of interest to find a point where $q_S$ achieves its minimum value.  Amenta et al.~\cite{AmeBerEpp-Algs-99} define {\em quasiconvex programming} as a formalization of this search for the minimum of $q_S$.  Linear programming can be seen as a special case of quasiconvex programming in which all the functions $q_i$ are linear.

More formally, Amenta et al. define a {\em nested convex family} to be a map
$\kappa(\lambda)$ from the nonnegative real numbers to compact convex sets in
$\R^d$ such that if
$\lambda_1<\lambda_2$ then
$\kappa(\lambda_1)\subset\kappa(\lambda_2)$, and such that
for all $\lambda$, $\kappa(\lambda)=\bigcap_{\lambda'>\lambda}\kappa(\lambda')$.
Any nested convex family $\kappa$ determines
a quasiconvex function $q_\kappa(w) = \inf\,\{\,\lambda \mathrel{|} w \in \kappa(\lambda)\,\}$
on $\R^d$, with level sets $q_\kappa^{\le\lambda}=\kappa(\lambda)$.
Conversely, when $q$ is quasiconvex, the closures of the level sets $q^{\le\lambda}$, restricted to a compact convex subdomain of $\R^d$, form a nested convex family.

Amenta et al. define a {\em quasiconvex program} to be formed by
a set of nested convex families
$S=\{\kappa_1,\kappa_2,\ldots \kappa_n\}$; the task to be solved is
finding the value
$$\lambda^*(S)=
\inf\Big\{\,
                (\lambda,w) \mathrel{\big|}
                                w\in \mathop{\textstyle\bigcap}\limits_{\kappa_i\in S}\kappa_i(\lambda)
\Big\}
$$
where the infimum is taken in the lexicographic ordering,
first by $\lambda$ and then by the coordinates of~$w$.

Quasiconvex programs can themselves be seen as a special case of
{\em generalized linear programs}.
These are optimization problems based on an objective function that maps sets to totally ordered values and that satisfies certain axioms~\cite{Ame-DCG-94,Gar-SJC-95,MatShaWel-TR-92};
quasiconvex programs are generalized linear programs because the function $\lambda^*(S)$ defined above satisfies these axioms~\cite{AmeBerEpp-Algs-99}.  One consequence of this generalized linear programming formulation is that the value of any quasiconvex program is determined by a {\em basis}, a subset of the nested convex families with cardinality $O(d)$.  Another consequence is that, when the dimension $d$ is bounded, quasiconvex programs can be solved by dual-simplex based randomized algorithms that perform a number of steps linear in the number of nested convex families in the input, where each step consists of solving a constant-sized subproblem.

As we now show, our problem of finding an optimal weight vector for our input recurrence can be expressed in this quasiconvex programming framework.

\begin{lemma}\label{lem:qc}
Let $c_{w,i}$ be as defined in the previous section.
Then the function $q_i(w)=c_{w,i}$ is quasiconvex.
\end{lemma}

\begin{proof}
We must show that $q_i^{\le\lambda}=\{w\mid c_{w,i}\le\lambda\}$ is convex.
Equivalently, expanding the definition of $c_{w,i}$ and using the monotonicity
of the function $r_{w,i}$, we must show convexity of the set
$$q_i^{\le\lambda}=\{w\mid r_{w,i}(\lambda)\ge 0\}
=\{w\mid\sum_j c^{-w\cdot\delta_{i,j}}\le 1\}.$$
But each summand of the sum in the right expression is an exponential of a linear function of $w$,
hence convex.  A sum of convex functions is convex, and its level set is a convex set.
\end{proof}

\begin{corollary}
We can find a pair $(c,w)$, where $c=\inf_w c_w$, $w\cdot t=1$, and
$w$ is a limit point of a sequence $w_\ell$ with $c_{w_\ell}$ converging to $c$,
as a solution to a quasiconvex program.
\end{corollary}

\begin{proof}
We form nested convex families from the closures of the level sets
of the functions $q_i(w)=c_{w,i}$.
The result follows from the definition of $c_w=\max c_{w,i}$.
\end{proof}

\begin{theorem}
If $(c,w)$ is found as the optimal solution to the quasiconvex program
defined as above for a recurrence for $F(x)$ with target vector $t$,
then $f(n)=F(n\,t)=O(c^n)$.
\end{theorem}

\begin{proof}
If $c_{w,i}=c<+\infty$ for all terms $i$, the result follows from Lemma~\ref{lem:wup}.
And if some term has $w\cdot\delta_{i,j}<0$, or has more than
one summands, then $c_{w,i}=+\infty$ but any sequence $w_\ell$ converging to $w$
has $c_{w_\ell}$ converging to $+\infty$, so the the upper bound is infinite and the result is vacuously true.  The only remaining case is that some terms have a single summand
$\delta_i$ with $w\cdot\delta_i=0$.  Such terms can not contribute to asymptotic growth of $F(n\,t)$ and the result follows by applying  Lemma~\ref{lem:wup} to the recurrence formed by omitting them from the definition of $F$.
\end{proof}

Therefore, we can use quasiconvex programming to find the weight vector $w$ yielding the tightest possible upper bound $O(c^n)$ on the asymptotic behavior of our recurrences.  Further, by finding a basis for the quasiconvex program, we can determine the recurrence terms that are critical for its asymptotics. The restriction $w\cdot t=1$ does not affect quasiconvexity, and in fact aids in the solution of our problem by reducing the effective dimension of the quasiconvex program to $d-1$.

\section{Implementation}

In this section we discuss two implementations of our optimization algorithms.  Our implementations use a numerical improvement technique which we call {\em smooth quasiconvex programming} and which will also feature in our later lower bound proof.

As stated earlier, generalized linear programming techniques can be used to solve our quasiconvex programs in $O(\tau)$ steps, where $\tau$ denotes the number of terms and each step consists of solving a constant sized subproblem.  Such a subproblem has only a constant number of potential bases, and we can test a basis by solving an algebraic system of equations
in the variables $c^{w[k]}$ where $w[k]$ denotes the $k$th coordinate of $w$.
However this approach appears likely to be cumbersome in practice.
Instead, we implemented a hill-climbing scheme for finding numerically the optimal value $(c,w)$ for our quasiconvex program and for certain other quasiconvex programs.

We have in mind two different purposes for an implementation of our analysis algorithm.
First, such an implementation could be used for exploratory analysis: computing a rough estimate of the running time of an algorithm, with enough accuracy to determine whether it improves other similar algorithms for the same problem, and determining the worst cases in an algorithm's case analysis, in order to refine that analysis to produce a better algorithm.  For this sort of application, it is important that the running time be fast enough to be usable at interactive or near-interactive speeds; an implementation using floating point arithmetic is appropriate.  Second, we would like to be able to publish guaranteed worst-case bounds on algorithms, for which approximate numeric schemes such as floating point that lack error bounds are inappropriate; in this setting the longer running times associated with exact or interval arithmetic may be acceptable.

With these considerations in mind, we implemented our algorithm twice, once using floating point and a second time using {\tt XR}~\cite{XR}, an exact real-number computation package for the Python programming language~\cite{Python}.  The results of the second implementation are guaranteed to be valid upper bounds for the recurrence in question.  In {\tt XR}, a real number $\rho$ is represented as a data structure that is capable of
constructing a multiprecision integer representing the rounded value of $2^b\rho$, for any integer $b$; this rounded value is required to be within unit distance of the true value.
We extended this package so that it could evaluate as exact real numbers the values $c_w$ given an input vector $w$ the coordinates of which are also real numbers, by performing an appropriate binary search using the values of the function $r_w(c)$.  We then implemented a quasiconvex programming algorithm to search for a sequence of vectors $w_\ell$ converging to the infimum $(c,w)$ of the quasiconvex program.  We describe in more detail below our exact arithmetic implementation; our floating point implementation is similar.

\begin{figure}[t]
\centering\includegraphics[width=4in]{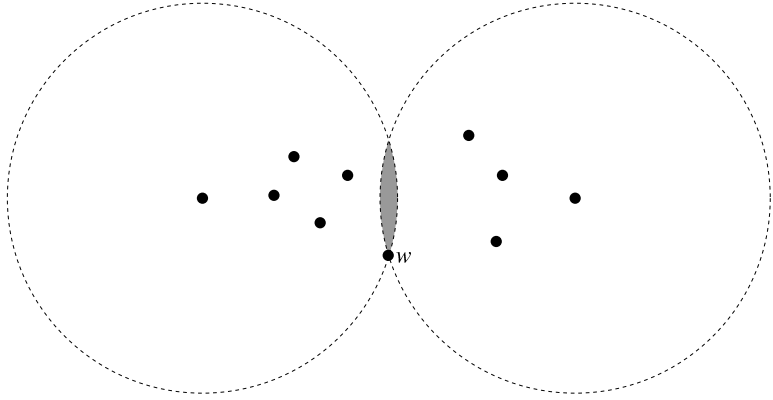}
\caption{Example showing the difficulty of applying standard gradient descent methods to quasiconvex programming.  The function to be minimized is the maximum distance to any point;
only points within the narrow shaded intersection of circles have function values smaller than the value at point~$w$.}
\label{fig:cr}
\end{figure}

If all functions $q_i(w)$ are quasiconvex, the function $q(w)=\max_i q_i(w)$ is itself quasiconvex, so we can apply hill-climbing procedures to find its infimum.  However, although in our application the individual functions $q_i$ are smooth, their maximum $q$ may not be smooth, so it is difficult to apply standard gradient descent techniques.  The difficulty may be seen, for instance, in a simpler quasiconvex programming problem: determining the circumradius of a planar point set (Figure~\ref{fig:cr}).  A basis for the circumradius problem may consist of either two or three points; the quasiconvex functions used to solve the problem are simply the distances (or squared distances) from each input point, and the function $q(w)$ measures the distance from $w$ to the farthest point.  But if a point set has only two points in its basis, and our hill climbing procedure for circumradius has reached a point $w$ equidistant from these two points and near but not on their midpoint, then improvements to the function value $q(w)$ may be found only by moving $w$ in a narrow range of directions towards the midpoint.  Standard gradient descent algorithms may have a difficult time finding such an improvement direction.  Similar behavior arises naturally in our recurrence analysis problem. For instance in a recurrence in~\cite{csds0302030} for the time bound of a TSP algorithm on cubic graphs, there are only two critical terms despite the problem being defined over a three-dimensional vector space, and these two terms lead to behavior very similar to the circumradius example.

To avoid these difficulties, we use the following algorithm, which we call {\em smooth quasiconvex programming}.  We assume we are given as input a set of quasiconvex real functions $q_i(w)$.
Further, we assume that for each $i$ we also can compute a vector-valued function $q_i^*(w)$,
satisfying the following properties:
\begin{enumerate}
\item If $q_i(x)<q_i(w)$, then $(x-w)\cdot q_i^*(w)>0$, and
\item If $q_i^*(w)\cdot v>0$, then for all sufficiently small $\epsilon>0$, $q_i(w + \epsilon v)< q_i(w)$.
\end{enumerate}
For the circumradius example, for instance, $q_i^*$ should be a vector pointing from $w$ towards the $i$th input point.
The requirements on $q_i^*$ can be described geometrically, as follows: we assume that the level set $q_i^{\le\lambda}$
is a {\em smooth} convex set, one that has at each of its boundary points a unique tangent plane.
The vector $q_i^*(x)$ is then an inward-pointing normal vector to the tangent plane
to $q_i^{\le q(x)}$ at $x$ (essentially it is just the negation of the gradient of $q$). The functions $q_i(w)$ arising in our recurrence analysis problem have this smoothness property,
and (as we discuss in more detail in the next section) the vectors $q_i^*(w)$ can be constructed by evaluating the partial derivatives for each of the coordinates of $x$ in the expression
$r_{x,i}(c_w)$ at $x=w$.

Our smooth quasiconvex programming algorithm then consists of selecting an initial value for $w$, and a desired output tolerance,
and repeating the following steps:
\begin{enumerate}
\item Compute the set of vectors $q_i^*(w)$,
for each $i$ such that $q_i(w)$ is within the desired tolerance of $\max_i q_i(w)$.
\item Find a vector $v$ such that $v\cdot q_i^*(w)>0$ for each vector $q_i^*(w)$ in the computed set.
If no such vector exists, $q(w)$ is within the tolerance of its optimal value and the algorithm terminates.
\item Search for a value $\epsilon$ for which $q(w+\epsilon v)\le q(w)$,
and replace $w$ by $w+\epsilon v$.
\end{enumerate}

Our actual implementation augments this procedure by an outer scaling loop that gradually decreases the tolerance,
so that multiple terms of the recurrence can influence the computation in step 1 even when the current value of $w$ is only a rough approximation to the true optimum.  We also terminate the loop when the improvement to $q(w)$ becomes much smaller than the tolerance, even when the termination condition of step 2 is not met, in order to handle situations in which the optimal basis is less than full-dimensional.

The search for a vector $v$ in step 2 can be expressed as a linear program.
However, when the dimension of the quasiconvex program is at most two
(equivalently, the number of variables in the recurrence to be solved is at most three)
it can be solved more simply by sorting the vectors $q_i^*(w)$ radially around the origin
and choosing $v$ to be the average of two extreme vectors.

In our implementation of step 3, we perform a doubling search for the largest $\epsilon$ leading to a smaller value of $q(w+\epsilon v)$, and then reduce the resulting $\epsilon$ by a factor of two before replacing $w$,
to attempt to control situations where the value $w$ oscillates around the true optimal value.

Both due to our use of exact real arithmetic, and due to the implementation in Python, a relatively slow interpreted language, our exact arithmetic implementation is not fast, taking several hours on a laptop computer to solve moderately sized 3-variable recurrences to 64 bits of precision.  However our floating point implementation is able to run at interactive speeds, taking roughly one or two seconds on a recent-model laptop to solve recurrences such as the one in Table~\ref{tbl:bigrec}.
We believe that significant improvements in runtime of our exact arithmetic implementation would be possible both by tuning the implementation and by using a faster software base.  However it is encouraging that, in the trials we attempted, the algorithm appears to exhibit linear convergence to the correct function value as well as to the optimal weight vector coordinates: the number of iterations of the algorithm appears to be proportional to the number of bits of precision desired.

\section{Lower Bounds}

In this section we prove that the upper bounds found by our optimal weighting technique are tight to within a polynomial factor.

\begin{figure}[t]
\centering
\includegraphics[width=5.5in]{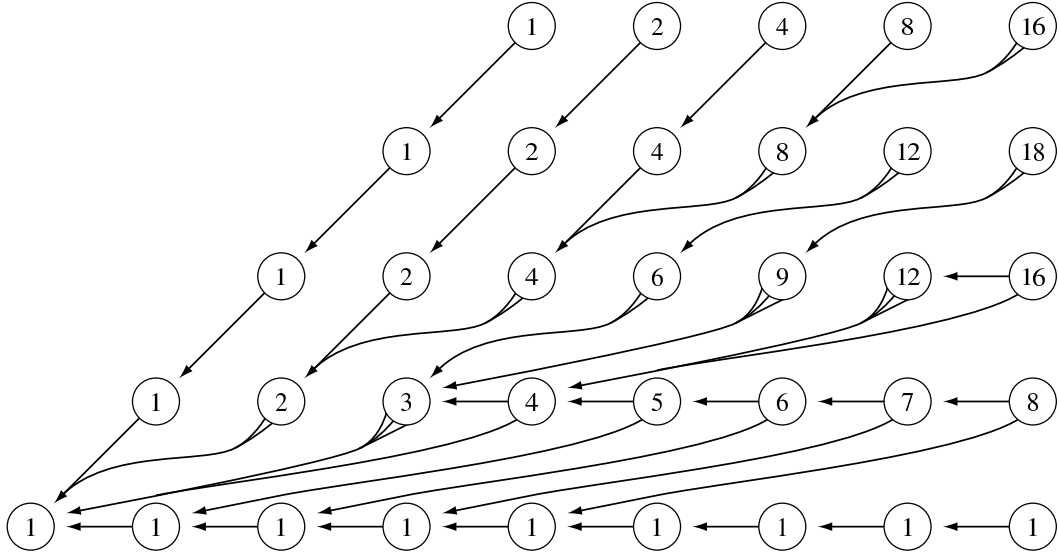}
\caption{A portion of the infinite graph $G_{i^*}$ for the recurrence $T(n,k)$ for listing
maximal independent sets of cardinality $k$ in an $n$-vertex graph, described in the introduction.
The horizontal position of a vertex indicates its $n$ coordinate and the vertical position indicates its $k$ coordinate.  To simplify the drawing, edges are shown in a confluent style~\cite{cs.CG/0212046} in which
multiple edges are allowed to merge before reaching their destination.
Each vertex is labeled with the number of paths in the graph from that vertex to the origin.}
\label{fig:gis}
\end{figure}

In order to find lower bounds for the asymptotic behavior of our recurrences, it is useful to have the following combinatorial interpretation of their values.  For any $x\in\Z^d$,
let $i^*(x)=\argmax_i \sum_j F(x - \delta_{i,j})$.  That is, $i^*$ is the index of the term in the recurrence that determines the value of $F(x)$.  If $\mu(x):\Z^d\mapsto\Z$
is any function mapping vectors to recurrence term indices, define an infinite graph $G_\mu$,
the vertices of which are vectors in $\Z^d$, with an edge from $x$ to $x-\delta_{i,j}$
whenever $i=\mu(x)$ and $\delta_{i,j}$ is a summand in term~$i$ of the recurrence.
Let $\pi_\mu(x)$ denote the number of paths from $x$ to zero in $G_\mu$.
It is not hard to show by induction that $\pi_{i^*}(x)=F(x)$, so $F$ can be interpreted as counting paths in a graph.  Moreover, for any $\mu$, $\pi_\mu(x)\le F(x)$.
An example of the graph $G_{i^*}$, for the recurrence discussed in the introduction, is shown in Figure~\ref{fig:gis}; each vertex $x$ in the figure is labeled with the number $\pi_{i^*}(x)=F(x)$.
We will find a lower bound for $F(x)$ by counting paths in a graph $G_\mu$,
where $\mu(x)$ will not necessarily equal $i^*(x)$.

Let $\lambda^*(S)=(c,w)$ be the optimal solution to the quasiconvex program formulated earlier for the upper bounds for our recurrence, and let $B$ be a {\em basis} of the quasiconvex program; that is, a minimal subset of terms of the recurrence such that the quasiconvex program formed from the subset has the same value $\lambda^*(B)=\lambda^*(S)$ as that for the whole recurrence.
For any term $i$ in $B$, and any summand $j$ in term $i$,
let $p_{i,j}=c^{-w\cdot\delta_{i,j}}$, and let $D_i=\sum p_{i,j}\delta_{i,j}$.

\begin{lemma}\label{lem:edgeprob}
For any term $i$ in $B$, $\sum_j p_{i,j}=1$.
\end{lemma}

\begin{proof}
If term $i$ has only one summand, then it belongs to $B$ exactly when $w\cdot\delta_{i,j}=0$ and $p_{i,j}=1$.  Otherwise, $q_i(w)$ is continuous and term $i$ can belong to $B$ only
when $q_i(w)=c$.  But then $r_w(c)=1-\sum p_{i,j}=0$ since $q_i$ is defined as having the value that makes this equation true, so $\sum p_{i,j}=0$.
\end{proof}

\begin{lemma}
The vector $D_i$ defined above is a scalar multiple of the vector $q_i^*$
defined for the smooth quasiconvex programming algorithm of the previous section.
\end{lemma}

\begin{proof}
The vector $q_i^*(w)$ is the vector such that $q_i(w+\epsilon v)<w_i(w)$ (for sufficiently small $\epsilon$) if and only if $v\cdot q_i^*>0$.  Expanding the definition of $q_i$, as the value $c$ such that $r_{w,i}(c)=0$, we see that $q_i^*(w)$ can equivalently be defined as a vector $v$ such that $r_{w+\epsilon v,i}(c)>0$ (for sufficiently small $\epsilon$) if and only if $v\cdot q_i^*>0$.Therefore, $q_i^*$ can be computed as the gradient of the vector function $\phi(x)=r_{x,i}(c)$, evaluated at $x=w$.
Expanding this gradient by computing partial derivatives for each component of its vector argument, we arrive at the definition of $D_i$.
\end{proof}

\begin{lemma}\label{lem:unbiased}
There exist values $b_i$, $0\le b_i\le 1$, so that $\sum b_i=1$
and so that $\sum b_i D_i$ is a scalar multiple of the target vector $t$.
\end{lemma}

\begin{proof}
The smooth quasiconvex programming algorithm of the previous section will find an improvement to solution $(c,w)$ whenever there exists $v$, perpendicular to $t$, having positive dot product with all the vectors $q_i^*(w)$.  The pair $(c,w)$ used to define $b_i$ and $D_i$ is assumed optimal, so can not be improved.  Therefore, $v$ does not exist and the origin must be contained in the convex hull of projections perpendicular to $t$ of the vectors $q_i^*$.  Any vector in the convex hull of a set of vectors can be expressed as a convex combination of those vectors, and the same convex combination (when viewed in $\R^d$ rather than its projection perpendicular to $t$) has the property stated in the lemma.
\end{proof}

We are now ready to describe the graph $G_\mu$ used in our lower bound construction.
For each $x\in Z^d$, we choose $\mu(x)$ to be one of the terms in the basis $B$,
independently at random, with probability $b_i$ for choosing term $i$.
We then let $G_\mu$ be the infinite graph formed from $\mu$ as described at the start of the section.  For each choice of $\mu$ the number of paths $\pi_\mu(n\,t)$ from $n\,t$ to the origin in $G_\mu$ forms a valid lower bound for the quantity $f(n)=F(n\,t)$ that we are trying to estimate,
so the expected number of paths (averaged over the choice of $\mu$) also forms a valid lower bound.

To count the expected number of paths in $G_\mu$, we use the following random walk process.
Start from the vertex $n\,t$, and then from any vertex $x$ choose randomly among the
edges leading away from $x$, independently for each $x$.  The set of edges at $x$ is in one-to-one correspondence with the summands of term $\mu(x)$, and we choose summand $j$ with probability $p_{\mu(x),j}$.  As shown in Lemma~\ref{lem:edgeprob} these probabilities add to one at each vertex.  We continue this random walk process until we reach a vertex $x$ with
$w\cdot x=0$.

\begin{lemma}\label{lem:equiprob}
If a path from $n\,t$ to the origin can be formed by the random walk described above, it has probability
$c^{-n}$ of being chosen.
\end{lemma}

\begin{proof}
More generally it is easy to show by induction on the length of the path that a path from $x$ to $y$ has probability $c^{w\cdot(y-x)}$ of being chosen.  The result follows from the choice of starting point $x=n\,t$ and the constraint that $w\cdot t=1$.
\end{proof}

\begin{lemma}\label{lem:origprob}
The random walk described above reaches the origin with
probability $\Omega(n^{(1-d)/2})$.
\end{lemma}

\begin{proof}
By Lemma~\ref{lem:unbiased}, the projections perpendicular to $t$ of the vertices of the path
form an unbiased random walk in $\R^{d-1}$, with $O(n)$ steps of constant size, and the result follows from the standard theory of such walks.
\end{proof}

\begin{theorem}
\label{thm:lb}
$f(n)=F(n\,t)=\Omega(c^n n^{(1-d)/2})$.
\end{theorem}

\begin{proof}
By Lemmas \ref{lem:equiprob} and~\ref{lem:origprob},
there is the given expected number of paths in $G_\mu$ from $n\,t$ to the origin.
The result follows from the facts that this number of paths is less than the
number of paths between the same two vertices in $G_{i^*}$ (since $i^*$ is defined to maximize the number of paths) and that the number of paths in $G_{i^*}$ is exactly $F(n\,t)$.
\end{proof}

\section{Conclusions and Open Problems}

We have shown that the solution to the recurrence
$$F(x) = \max_i \sum_j F(x - \delta_{i,j}),$$
for $x=n\,t$
may be upper and lower bounded within a polynomial factor of $c^n$,
where the number $c$ can be computed as the solution to a quasiconvex program
defined from the recurrence and the target vector~$t$.

It would be of interest to tighten these bounds: under what conditions can we determine the correct polynomial adjustment to the bound $c^n$?  It is consistent with our observations so far that
$F(n\,t)=\Theta(c^n n^{(|B|-d)/2})$ where $|B|$ is the cardinality of a basis for the quasiconvex program.  For instance this would fit the central binomial coefficients, with $|B|=1$ and $d=2$, as well as the recurrence used as an example at the start of this paper with $|B|=d=2$.  However there is too little evidence yet to state such a formula as a conjecture.

More generally, the work here is only a first step towards the automation of backtracking algorithm design and analysis.  Perhaps it would also be possible to automatically perform some of the case analysis used to design backtracking algorithms, and to determine the appropriate variables to use in setting up the recurrences used to analyze those algorithms, before automatically solving those recurrences, at least for simple constraint satisfaction type problems.  It would also be of interest to find ways of specifying algorithms of this type in such a way that their correctness can be proven automatically, especially in situations where repeated refinement based on our analysis tools has led to highly complex case analysis such as that appearing in Table~\ref{tbl:bigrec}.
Also, while we can find tight worst-case bounds on the solution of the recurrence derived from an algorithm, it may not always
be possible to construct an instance causing the algorithm itself to have that worst case time bound; it would be useful to determine conditions under which this recurrence-based analysis is tight.

In another direction, the proof of Theorem~\ref{thm:lb} hints at a theory of duality for quasiconvex programs that it would be of interest to explore.

\section*{Acknowledgements}

This research was supported in part by NSF grant CCR-9912338.
I would like to thank Jesper Byskov and George Lueker for helpful discussions and comments on drafts of this paper, and Keith Briggs for help with programming using~{\tt XR}.
 
\raggedright
\bibliographystyle{abuser}
\bibliography{qaba}
 \end{document}